\documentclass[reprint,aps,prb,superscriptaddress,amsmath,amssymb,floatfix,footinbib,longbibliography]{revtex4-1}
\usepackage{graphicx}
\usepackage{xcolor}
\usepackage{bm}
\usepackage[colorlinks, linkcolor= blue, citecolor = blue, urlcolor=blue]{hyperref}
\usepackage{float}
\usepackage[stable]{footmisc}

\newcommand{\ignoreY}[1]{\iffalse #1 \fi}

\def\bea{\begin{eqnarray}}
\def\eea{\end{eqnarray}}
\def\be{\begin{equation}}
\def\ee{\end{equation}}

\def\bal{\begin{aligned}}
\def\eal{\end{aligned}}

\begin{document}
\title{Quantum metric induced nonlinear thermal noise in ${\cal P} \cal{T}$ symmetric antiferromagnets}
	\author{Dibyanandan Bhowmick}
	\affiliation{Department of Physics, Indian Institute of Technology, Kanpur-208016, India.}
    \author{Amit Agarwal}
    \email{amitag@iitk.ac.in}
    \affiliation{Department of Physics, Indian Institute of Technology, Kanpur-208016, India.}

\begin{abstract}
    Electrical current fluctuations provide a powerful and unconventional probe of band geometry in quantum materials. In particular, intrinsic noise components that are independent of the relaxation time reveal universal band-geometric properties of Bloch electrons. Here, we identify a distinct intrinsic contribution to current noise at second order in the electric field, which is governed by the quantum metric. This effect arises from field-induced modifications to Bloch wavefunctions and band energies, and it dominates in ${\cal P}{\cal T}$-symmetric antiferromagnets where Berry curvature–based contributions are forbidden by symmetry. Applying our theory to CuMnAs, we demonstrate that thermal noise provides a direct signature of quantum metric in ${\cal P}{\cal T}$-symmetric antiferromagnets.
\end{abstract}

\maketitle

    \section{Introduction} 
    Band-geometric quantities such as Berry curvature play a central role in characterizing topological phases and explaining a wide range of transport and optical responses in quantum materials \cite{Chang_96_SemClassDynam, 
    Sundaram_Niu_99, Xiao_2010_RMP_Berry_Phase_Electronic}. The Berry curvature offers important insights for understanding the flow of charge, spin, valley, and heat currents in both linear and nonlinear regimes \cite{ 
    Culcer_03_AHE_Param, 
    Sinitsyn_07_AHE_SemiClassics,  
    Nagaosa_10_AHE_RMP, 
    Chang_2013_Exp_QAHE, 
    Sodemann_2015_NL_BCD_Hall, 
    Liu_17_AnnRev_QAHE_ThAndExp, 
    Xu_2018_ExpBCD_WTe2, 
    Gao_19_NL_Current_SemDyn, 
    Xiao_20_BC_Memory, 
    Ho_21_Hall_Corrugated_graphene, 
    Kumar_21_Exp_RoomTemp_NLHall_TaIrTe4, 
    He_21_Exp_FreqDoubling_Bi2Se3, 
    Lahiri_22_BC_NL_magnetoresistance,
    Sinha_2022_BCD_top_tran_moire, 
    Chakraborty_22_NLAHE_Top_trans_TDBLG,
    Kamal_22_SHG,
    Bhalla_2023_symm, 
    Chang_23_AHE_RMP, 
    Adak_24_Moire_Berry,
    Sankar_24_BC_Exp_Quadra,
    Datta_2024_Moire_Fermi_Surface_Malleability,
    Layek_25_QGT_Moire_Stacking,
    Varshney_25_NL_Nernst_Seebeck,
    Ahmed_2025_Lifshitz, 
    Murakami_03_spin_current, 
    Bernevig_06_ThQSHE_HgTe, 
    Markus_07_ExpQSHE_HgTe, 
    Sinova_15_SHE_RMP, 
    Xiao_07_Lin_Vlley, 
    Ghorai_24_NVHE, 
    Debottam_20_Nernst_thermal, 
    Harsh_23_Thermal}. Beyond linear transport regime, another gauge invariant band-geometric quantity, the quantum metric tensor (QMT), arises naturally in nonlinear conductivity \cite{Provost_80_QMT, Liu_21_BCP_Compens_AFM, 
    Wang_21_NLHE_CuMnAs,
    Gao_Barun_2023_QMT_NL_Hall, 
    Wang_2023_QMT_NL_antiferromag, 
    Harsh_23_Magnon, 
    Shibalik_23_cond_BCP, 
    Kaplan_2024_01,
    Kamal_25_NL_QMT_Surface_dominated,
    Varshney_25_NL_Nernst_Seebeck}.  
    Given the success of band-geometric quantities in explaining charge transport responses, it is natural to seek alternative experimental probes that capture the impact of band-geometric quantities. This motivates our search for signatures of band-geometric quantities in thermal fluctuations of linear and non-linear longitudinal and Hall currents. 
    
    In repeated measurements, the average value (mean) and the variance (fluctuation) are key observables. The latter captures the noise and has attracted some interest from theoretical \cite{Blanter_2000_Shot_Noise, Neupart_2013_Noise, Aseev_16_th_Shot_Noise_2DTI, Vayrynen_2017_th_MagMomt_Noise, Kurilovich_19_eBunching_Helical_edge, Safi_20_FlucDiss_Strong_OE, Probst_22_SubToSuper_Poisson, Miaomiao_2023, Xiang_2024_noise_photocurrent, Kruchkov_2024_Gap_by_Noise} and experimental \cite{Piatrusha_17_exp_Shot_Noise_HgTe, Stevens_2019_Noise_InAs/Ga(In)Sb_Corbino} perspectives. While the role of band-geometric quantities in transport and optical properties has been well explored, their signatures in quantum and thermal fluctuations (or in shot noise and thermal noise) are relatively less explored. The impact of Berry curvature-induced anomalous Hall velocity on the thermal noise was recently investigated \cite{Miaomiao_2023} in both linear and non-linear response regimes. Interestingly, the Berry curvature itself vanishes in ${\cal P} \cal{T}$ symmetric antiferromagnets. Consequently, the corresponding linear (anomalous Hall) and nonlinear (Berry curvature dipole) currents vanish. However, the Berry Connection Polarizability (BCP) has been shown to induce intrinsic longitudinal and Hall conductivity in the nonlinear response regime \cite{Liu_21_BCP_Compens_AFM, Wang_21_NLHE_CuMnAs, Wang_2023_QMT_NL_antiferromag, Gao_Barun_2023_QMT_NL_Hall, Shibalik_23_cond_BCP, Bhalla_2023_symm}. This motivates a natural question, `What are the signatures of Berry curvature polarizability in nonlinear thermal noise measurements?'

    Here, we address this question in detail and demonstrate the impact of BCP on thermal noise in noncentrosymmetric antiferromagnets. In Sec. II and III, we first discuss the general theory for calculating thermal noise from the expression of current and revisit the earlier results\cite{Miaomiao_2023}. Then in Sec. IV, we calculate the quantum metric-induced contributions to thermal noise after incorporating perturbative corrections \cite{Gao_2014, Liu_2022_3rd_Hall, Huang_2023_NL_PHE} in wavefunction and band energy in the presence of an external DC electric field. This enables us to obtain the band-normalized metric (i.e., BCP) regulated thermal noise at second-order of the electric field. We illustrate our findings using both the tilted Dirac model and the antiferromagnet CuMnAs.. 
    
\section{Current Operator and Thermal Noise}
To show that intrinsic current noise captures band-geometric information, we first describe the electric-current operator and its equal-time noise tensor.

Consider the Bloch state of $n$th band at crystal momentum $\hbar \mathbf{k}$,  $|n(\mathbf{k})\rangle$.  In the second quantization framework, the current operator along $a$ direction is given by 
\begin{equation}\label{eq:J_operator}
  \hat{J}_{a} 
  =- e \sum_{m,n,\mathbf{k}}
  V^{m n}_{a \mathbf{k}} 
  \hat{c}^{\dagger}_{m\mathbf{k}}\hat{c}_{n\mathbf{k}}~, 
\end{equation}
with band indices $n, m$, and Cartesian directions denoted by $a$. The  velocity matrix elements are given by 
\begin{equation}
V^{m n}_{a\mathbf{k}} 
      = \langle m(\mathbf{k}) 
      \left| \frac{1}{\hbar} \frac{\partial \hat{H}(\mathbf{k})}{\partial k_a} \right| n (\mathbf{k})\rangle~.
\end{equation} 
The fermionic operators satisfy
\(\langle\hat{c}^{\dagger}_{m\mathbf{k}}\hat{c}_{n\mathbf{k}}\rangle
     =f_{n\mathbf{k}}\delta_{mn}\),
where $f_{n\mathbf{k}}$ is the non-equilibrium distribution function. It reduces to the Fermi-Dirac distribution function in equilibrium. 

Noise originates from fluctuations around the mean current. Fluctuations about the mean are captured by 
\(\Delta\hat{J}_{a}\equiv\hat{J}_{a}-\langle\hat{J}_{a}\rangle\),
for which \(\langle\Delta\hat{J}_{a}\rangle=0\).
The equal-time, symmetrised noise tensor is defined as 
\begin{equation}\label{eq:noise_def}
  \eta_{ab}=\frac{1}{2}
  \bigl\langle
    \Delta\hat{J}_{a}\Delta\hat{J}_{b}
   +\Delta\hat{J}_{b}\Delta\hat{J}_{a}
  \bigr\rangle~.
\end{equation}
Here, $\eta_{aa}$ is the current variance or auto-correlation and $\eta_{ab}$ $(a\neq b)$ measures cross-correlations.  The symmetrisation ensures that $\eta_{ab}$ is real even when $\hat{J}_{a}$ and $\hat{J}_{b}$ do not commute. Although Eq.~\eqref{eq:noise_def} can be extended to unequal times, the equal-time form suffices for the steady-state, intrinsic noise that we focus on.  Below, we show that $\eta_{ab}$ provides a disorder-independent probe of the band-geometric quantities such as the quantum metric and the Berry curvature. 

To calculate the current noise, we substitute the second-quantized current operator, Eq.~\eqref{eq:J_operator}, into the symmetrized definition of noise, Eq.~\eqref{eq:noise_def}, and decompose the result into interband ($m \neq n$) and intraband ($m = n$) sectors. The interband contribution survives down to zero temperature and is therefore called {\it shot noise}. The intraband contributions vanish as $T\to 0$ and these are referred to as {\it thermal noise}.

The shot noise is given by \cite{Miaomiao_2023}, 
\begin{equation}\label{eq:shot_noise}
  \eta^{\text{shot}}_{ab}=
  e^{2}
  \sum_{\mathbf{k}}\!
  \sum_{n\neq m}
  \sum_{m}
  f_{mn\mathbf{k}}\;
  V^{n m}_{a\mathbf{k}}\,
  V^{m n}_{b\mathbf{k}}~. 
\end{equation}
Here, we have defined the interband prefactor, 
\[
  f_{mn\mathbf{k}}
  =\frac12\!\left[
    f_{m\mathbf{k}}\bigl(1-f_{n\mathbf{k}}\bigr)
   +f_{n\mathbf{k}}\bigl(1-f_{m\mathbf{k}}\bigr)
  \right].
\]
Since $f_{mn\mathbf{k}}$ involves both occupied and empty states\cite{Miaomiao_2023, Xiang_2024_noise_photocurrent}, $\eta^{\text{shot}}_{ab}$ remains finite even at $T=0$.  It dominates the total noise whenever the field-induced energy drop $eEL$ across the sample exceeds the thermal energy scale, i.e.,\ for $k_BT\!\ll\!eEL$ \cite{Miaomiao_2023}. 

The intraband thermal-noise contribution is given by 
\begin{equation} 
\eta^{\text{th}}_{ab} 
= e^{2}
\sum_{n,\mathbf{k}}
f_{n\mathbf{k}}\bigl(1-f_{n\mathbf{k}}\bigr)\,
V^{nn}_{a\mathbf{k}}\,
V^{nn}_{b\mathbf{k}}~.
\label{eq:thermal_noise}
\end{equation}
Note that the function 
\[
f_{n\mathbf{k}} ^{eq}
\bigl(1-f_{n\mathbf{k}} ^{eq} \bigr)
= -\,k_{\mathrm B}T\,
\frac{\partial f_{n\mathbf{k}} ^{eq} }{\partial\varepsilon_{n\mathbf{k}}}~.
\]
Due to this, we have $\eta^{\text{th}}_{ab}$ vanishing as $T\!\to\!0$.  This contribution becomes the dominant source of current noise when the thermal energy exceeds the field-induced potential drop across the sample, i.e., for $k_{\mathrm B}T\gg eEL$\,\cite{Miaomiao_2023}. We will focus on this contribution in the rest of the paper. 

To access different band-geometric contributions to the thermal noise, we expand it in powers of the static electric field $E$. Focusing on linear and second-order contributions, we have 
\begin{equation}
\eta^{\text{th}}_{ab}= 
\Gamma^{(1)}_{ab}
+\Gamma^{(2)}_{ab}
+\mathcal{O}(E^{3})~,
\end{equation}
with $\Gamma^{(n)}_{ab}\propto|E|^{\,n}$. Both linear and second-order contributions decompose naturally into \emph{extrinsic} parts which scale with various powers of the relaxation time $\tau$ and an \emph{intrinsic} part that is $\tau$-independent and has a purely band-geometric origin. In Sec. III we revisit \cite{Miaomiao_2023} the calculation of $\Gamma^{(1)}_{ab}$ and $\Gamma^{(2)}_{ab}$, showing that their intrinsic parts are governed by the Berry Curvature. In Sec. IV, we develop the theory of the thermal noise governed by the quantum metric, which is the main contribution of this work.

\section{Semiclassical Formalism}

To evaluate the linear \(\Gamma^{(1)}_{ab}\) and quadratic \(\Gamma^{(2)}_{ab}\) field contributions to the thermal-noise tensor, we require explicit expressions for the diagonal velocity matrix elements and the nonequilibrium distribution function.  For a uniform dc electric field \(\mathbf E\) in the absence of a magnetic field, the semiclassical equations of motion~\cite{Sundaram_Niu_99} give the diagonal velocity of band \(n\) to first order in \(\mathbf E\). We have,  
\begin{equation}
\label{eq:v_diagonal}
V^{nn}_{a\mathbf k}
=\frac{1}{\hbar}\frac{\partial \varepsilon_{n\mathbf k}}{\partial k_a}
-\frac{e}{\hbar}\,\epsilon_{abc}\,
\Omega^{n}_{b\mathbf k}\,E_c~,
\end{equation}
where \(\varepsilon_{n\mathbf k}\) is the band dispersion. In Eq.~\eqref{eq:v_diagonal}, the first term is the band velocity ($ v_{a, \mathbf{k} } ^n $), and the second term is the anomalous Hall velocity. The Berry curvature appearing in Eq.~\eqref{eq:v_diagonal} is expressed as the curl of the Berry connection as,  
\begin{equation}
\label{eq:berry}
\Omega^{n}_{a\mathbf k}
=\epsilon_{abc}\,\partial_{k_b}A^{n n}_{c\mathbf k}, 
\qquad 
A^{n n}_{c\mathbf k}
=\bigl\langle n(\mathbf k)\,|\,i\partial_{k_c}n(\mathbf k)\bigr\rangle~ .
\end{equation}

To obtain the nonequilibrium distribution function, we use the semiclassical Boltzmann transport equation in the relaxation-time approximation. It is given by, 
\begin{equation}
\label{eq:boltz}
\frac{\partial f_{n\mathbf k}}{\partial t}
+\dot{\mathbf r}\!\cdot\!\nabla_{\mathbf r}f_{n\mathbf k}
+\dot{\mathbf k}\!\cdot\!\nabla_{\mathbf k}f_{n\mathbf k}
=\frac{f^{eq}_{n\mathbf k}-f_{n\mathbf k}}{\tau}~.
\end{equation}
In a spatially uniform system in the steady state, it reduces to 
\(\dot k_a\,\partial_{k_a}f_{n\mathbf k}
=(f^{eq}_{n\mathbf k}-f_{n\mathbf k})/\tau\). 
The rate change of crystal momentum\cite{Sundaram_Niu_99} is given by \(\dot k_a=-eE_a/\hbar\), assuming that there is no magnetic field. Using a perturbative expansion of the distribution function in powers of the electric field, 
$ f_{ n \mathbf{k} }  
= f^{eq} _{ n \mathbf{k} }  
+ f^{(1)} _{ n \mathbf{k} }  
+ f^{(2)} _{ n \mathbf{k} }  
+ \cdots $, we obtain the recursive solution, 
\begin{equation}
\label{eq:f_series}
f^{(\ell)}_{n\mathbf k}
=\Bigl(\frac{e\tau}{\hbar}\Bigr)^{\!\ell}
\,(E_a\partial_{k_a})^{\ell}f^{eq}_{n\mathbf k}~.
\end{equation}
Here, $ f^{eq} _{ n \mathbf{k} }  $ is the equilibrium Fermi-Dirac distribution function, given by
 \begin{equation} \label{eq:Fermi-Dirac-Distribution}
    f^{eq} _{ n \mathbf{k} } 
    = \Big[ 
    \exp \Big(
    \frac{ \varepsilon_{ n \mathbf{k} } - \mu }{ k_B T }
    \Big) 
    + 1 
    \Big]^{-1}~,
    \end{equation}
with $T$ being the temperature and $\mu$ denoting the chemical potential of the sample. 
    
    Following Ref.~[\onlinecite{Miaomiao_2023}], the thermal noise linear in the electric field consists of both the intrinsic and the extrinsic components. The extrinsic contribution is $\tau$-linear, and it is given by, 
    \begin{equation} \label{eq:linear_noise_extrinsic}
        \Gamma^{(1)}_{a b} ~\Big|_{\text{ext}}
        = e^3 \tau
        \sum_{n, \mathbf{k} } ( 1 - 2 f_{ n \mathbf{k} } ^{eq} ) 
        \Big(
        \frac{ \partial f_{n \mathbf{k} } }{ \partial \varepsilon_{ n \mathbf{k} } }
        \Big)
        v_{ a \mathbf{k} } ^n 
        v_{ b \mathbf{k} } ^n
        v_{ c \mathbf{k} } ^n
        E_c ~~.
    \end{equation}
    This extrinsic component vanishes for time-reversal symmetric systems. What survives in ${\cal T}$-symmetric systems is the $\tau$-independent thermal noise. Arising from the unperturbed equilibrium distribution and the anomalous velocity term, intrinsic contribution is given by, 
    \begin{equation} \label{eq:linear_noise_intrinsic}
        \begin{split}
             \Gamma^{(1)}_{a b}\Big|_{\text{int}} & 
             = - \frac{e^3}{\hbar} k_B T 
            \sum_{n, \mathbf{k} } 
            f_{n, \mathbf{k} } ^{eq}
            \left[
            \partial_{k_a} ( \mathbf{\Omega} _{ \mathbf{k} } ^{n} \times \mathbf{E} )_b 
            + \partial_{k_b}  ( \mathbf{\Omega}_{ \mathbf{k} } ^{n} \times \mathbf{E} )_a\right] \\
            & =  - \frac{e^3}{\hbar} k_B T 
            \Big[
            ( \mathbf{D}_a \times \mathbf{E} )_b 
            + ( \mathbf{D}_b \times \mathbf{E} )_a 
            \Big] ~.
        \end{split}
    \end{equation}
    In the above equation, $ \mathbf{D}_a $ is the gauge-invariant Berry Curvature dipole (BCD), defined as $ \mathbf{D}_a 
    = \sum_{n, \mathbf{k} } f_{n \mathbf{k} } ^{eq} 
    ( \partial_{k_a} \mathbf{\Omega}_{ \mathbf{k} } ^{n} )  $.  
    The BCD is a $2^{nd}$ rank pseudo-tensor, capturing the band-geometric contribution in noise. 
    
    In contrast, the second-order noise contains terms proportional to both $\tau^2$ and $\tau$, along with an an intrinsic ($\tau^0$) component, $\Gamma_{a b }^{(2)} = \tilde\Gamma_{ a b }^{\rm quad}(\tau^2) + \tilde\Gamma_{ a b }^{\rm lin}(\tau) + \tilde\Gamma_{ a b }^{\rm int}(\tau^0)$~. The calculations reproduce the results of the $\tau^2$ component\cite{Miaomiao_2023},
    \begin{equation} \label{eq:quad_noise_tau2}
          \tilde\Gamma_{ a b }^{\rm quad} 
          =  \tau^2 \frac{e^4}{\hbar^2} 
          \sum_{n, \mathbf{k} }
           \Big[ \frac{\partial^2 g_2}{ \partial k_c \partial k_d } 
           + ( \partial_{k_c} f_{n \mathbf{k} } ^{eq} ) 
             ( \partial_{k_d} f_{n \mathbf{k} } ^{eq} ) 
           \Big] 
           v_{a \mathbf{k} } ^{n}  
           v_{b \mathbf{k} } ^{n} 
           E_c E_d~.
    \end{equation} 
    This is the Drude contribution to thermal noise. The derivatives of the Fermi-Dirac distribution function, along with the term containing 
    $ g_2 
    = f_{n \mathbf{k} } ^{eq} 
    ( 1 - f_{n \mathbf{k} } ^{eq} ) 
    = -k_B T 
    (\partial f_{n \mathbf{k} } ^{eq} / \partial \varepsilon_{ n \mathbf{k} } ) $ %
    indicate that this term captures the Fermi surface effect, and it vanishes if the Fermi energy lies in the bandgap.  

    The $\tau$-linear contribution is given by \cite{Miaomiao_2023}, 
    \begin{equation} \label{eq:quad_noise_tau1}
        \begin{split}
        \tilde\Gamma_{ a b }^{\rm lin} = 
        & ~~\tau \frac{e^4}{\hbar^2}
        \sum_{n, \mathbf{k} }
        ( 1 - 2 f_{n \mathbf{k} } ^{eq} ) 
        ( \mathbf{E} 
        \cdot \mathbf{\nabla}_{ \mathbf{k} } 
        f_{n \mathbf{k} } ^{eq} 
        )  \\ 
        & \times \Big[
           v_{a \mathbf{k} } ^{n} 
           ( \mathbf{\Omega}_{ \mathbf{k} } ^{n} \times \mathbf{E} )_b 
           + v_{b \mathbf{k} } ^{n} 
           ( \mathbf{\Omega}_{ \mathbf{k} } ^{n} 
           \times \mathbf{E} )_a
           \Big]~.
        \end{split}
    \end{equation}
    The term is linear in the relaxation time. It is a hybrid contribution involving both the band velocity and the band geometry. This term vanishes if time-reversal symmetry is present in a system.

    The intinsic term, independent of $\tau$ ($\propto \tau^0$) is given by, 
    \begin{equation} \label{eq:quad_noise_tau0}
        \tilde\Gamma_{ a b }^{\rm int} 
        =  k_B T \frac{e^4}{\hbar^2} 
        ~ \mathcal{E}_a ^T \Omega^{(2)} _{a b} \mathcal{E}_b ~~.
    \end{equation}
    Here, $a$ and $b$ are not summed over. This term, encoding the intrinsic contribution, involves the vector 
    $\mathcal{E}_a 
    = \mathbf{E} \times \hat{\mathbf{a}}$, 
    written as a column matrix. Here, the matrix $\Omega^{(2)} _{ a b}$, is the {\it fluctuation of the quantum geometric tensor}. It represents another band-geometric contribution to second-order thermal noise \cite{Miaomiao_2023} and is defined as,  
    \begin{equation} \label{eq:fluctuation_QGT}
        \Omega^{(2)} _{a b} 
        = \sum_{n, \mathbf{k} } 
        \Big( 
        - \frac{\partial f_{n \mathbf{k} } ^{eq} }{ \partial \varepsilon_{n \mathbf{k} } } 
        \Big) 
        \Omega^{n} _{ a \mathbf{k} } 
        \Omega^{n} _{ b \mathbf{k} }~.
    \end{equation}
    
\section{Berry-Connection Polarizability Contribution to Thermal Noise}

The results presented in previous section on intrinsic contributions to current noise, rely on the presence of finite Berry curvature in the system. These are  inapplicable to $\mathcal{PT}$-symmetric antiferromagnetic systems, where the Berry curvature vanishes identically for all points in the Brillouin zone. To uncover band-geometric effects beyond the Berry curvature, we extend the formalism by including electric-field corrections to both the Bloch states and the band dispersions. This reveals a novel intrinsic noise contribution governed by the quantum metric~\cite{Provost_80_QMT}, which emerges as the leading second-order intrinsic contribution in $\mathcal{PT}$-symmetric antiferromagnets such as CuMnAs. 

\begin{figure*}[t!]
        \includegraphics[width=.8\linewidth]{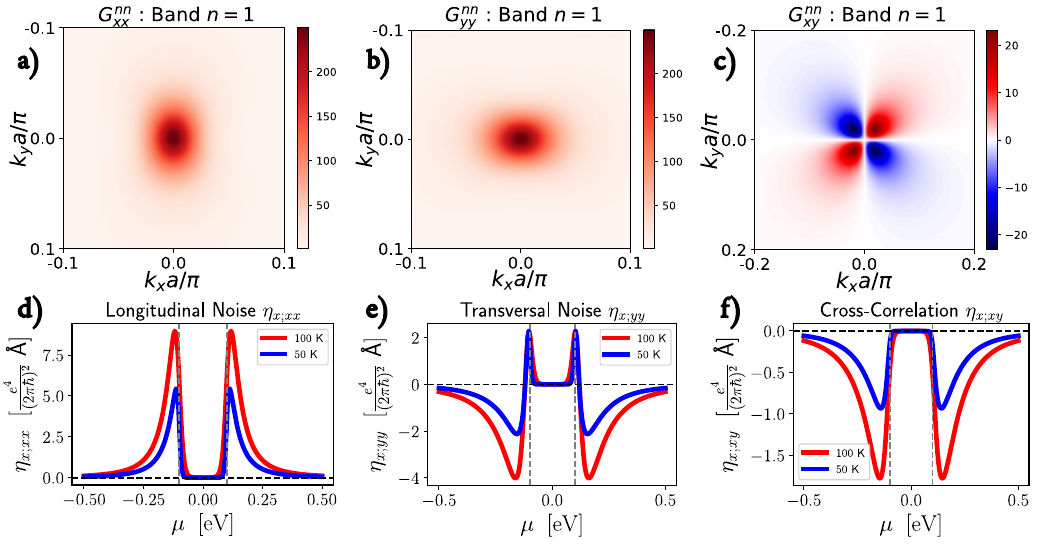}
        \caption{\textbf{Tilted Dirac model for surface states:} (a)–(c) BCP tensor components $G_{ab}^{nn}$ for the conduction band ($n=1$), shown in units of eV$^{-1}$. These components peak near the band edges $ (k_x, k_y) = (0,0) $ point. (d)–(f) Thermal noise $\eta_{x;ab}$ as a function of chemical potential $\mu$, with the electric field $\mathbf{E}=(1,0,0)$ V/m applied along the crystal $X$ axis. These correlations are suppresed inside the gap as they are Fermi surface effects. Additionally, these the noise components are peaked near the edges, reflecting their band-geometric origin. We have used $\hbar v_f = 1$ eV/Å, $\hbar v_t = 0.1$ eV/Å, and $\hbar \Delta = 0.1$ eV.
        \label{fig:Dirac}}
    \end{figure*}
    
An external dc electric field $\mathbf{E}$ induces a polarization of the intraband Berry connection~\cite{Gao_2014}, which is given by
\be 
A^{nn}_{a\mathbf{k}} \to A^{nn}_{a\mathbf{k}} + e G^{nn}_{a b, \mathbf{k}} E_b + \mathcal{O}(E^2)~, 
\ee
up to linear order in the field. Here, the factor $G^{nn}_{ab,\mathbf{k}}$ is the Berry-connection polarizability (BCP), a second rank real-valued tensor, which captures the quantum-metric contribution to the Berry connection. In terms of the band energies ($ \varepsilon_{n \mathbf{k} } $) and the interband ($\ell \neq n$) Berry connections $A^{n\ell}_{a\mathbf{k}} 
= \langle n(\mathbf{k}) | i \partial_{k_a} |\ell(\mathbf{k}) \rangle$, the BCP for a given band is given by~\cite{Gao_2014}
\be \label{eq:bcp_definition}
G^{nn}_{ab,\mathbf{k}} 
= 2 \Re \left[ 
\sum_{\ell \neq n} 
\frac{ A^{n\ell}_{ a,\mathbf{k} } A^{\ell n}_{ b,\mathbf{k} } }{\varepsilon_{ n\mathbf{k} } - \varepsilon_{ \ell\mathbf{k} } }
\right] ~.
\ee 
The BCP involves the band-resolved quantum metric $\mathcal{G}^{n \ell}_{ab,\mathbf{k}} 
= \Re \{ A^{n\ell}_{a\mathbf{k}} A^{\ell n}_{b\mathbf{k}} \}$, normalized by the energy difference between the bands, and summed over intermediate bands. 

This field-induced polarization of the Berry connection modifies the Berry curvature and the anomalous velocity. To the second order in the electric field, the anomalous velocity becomes \cite{Wang_21_NLHE_CuMnAs, 
Liu_21_BCP_Compens_AFM}
\begin{equation}
    \begin{split}
        v^{n, \text{anom}} _{a, \mathbf{k} } 
         = & -\frac{e}{\hbar} 
        \epsilon_{abc} \Omega^{nn} _{b,\mathbf{k} } E_c \\
        & + \frac{e^2}{ 2 \hbar }
        \Big(
        2 \partial_{k_a} G^{nn} _{ cd, \mathbf{k} }
        - \partial_{k_c} G^{nn} _{ ad, \mathbf{k} }
        - \partial_{k_d} G^{nn} _{ ca, \mathbf{k} }
        \Big)
        E_c E_d~. 
    \end{split}
\end{equation}
In addition to the anomalous velocity, the dc electric field also renormalizes the band energies~\cite{Liu_2022_3rd_Hall, Huang_2023_NL_PHE}, leading to, 
\be 
\varepsilon_{n\mathbf{k}} 
\to \varepsilon_{n\mathbf{k}} 
- \frac{e^2}{2} G^{nn}_{ab,\mathbf{k}} E_a E_b 
+ \mathcal{O}(E^3)~.
\ee 
This energy renormalization, in turn, introduces an electric field correction in the band velocity, 
\be 
 v^{n}_{a,\mathbf{k}} 
 = \frac{1}{\hbar} \partial_{k_a} \varepsilon_{n\mathbf{k}} 
 - \frac{e^2}{2\hbar} \left( \partial_{k_a} G^{nn}_{cd,\mathbf{k}} \right) E_c E_d 
 + \mathcal{O}(E^3)~.
 \ee 
The energy correction of the bands also leads to a modification in the equilibrium Fermi-Dirac distribution function,
\be 
f^{eq}_{n\mathbf{k}} \to f^{eq}_{n\mathbf{k}} 
+ \frac{e^2}{2} 
\left( 
\frac{\partial f^{eq}_{n\mathbf{k}}}{\partial \varepsilon_{n\mathbf{k}}} 
\right) 
G^{nn}_{ab,\mathbf{k}} E_a E_b 
+ \mathcal{O}(E^3) ~.
\ee
Since our analysis pertains to steady-state transport under a dc field, the carriers equilibrate to these field-renormalized bands. Consequently, we incorporate electric-field corrections consistently across the Bloch wavefunctions, band energies, and distribution functions.

Substituting these field-corrected expressions into the thermal-noise expression in Eq.~\eqref{eq:thermal_noise}, and retaining terms up to second order in $\mathbf{E}$, we obtain the BCP-induced intrinsic contribution to the thermal noise. It is specified by,  
\begin{widetext}
\begin{equation} \label{eq:bcp_noise}
\eta^{\text{th, int, BCP}}_{ab} 
= \frac{e^4 k_B T}{2} E_c E_d 
\sum_{n,\mathbf{k}} 
\left[ 
\frac{1}{\hbar} 
\left( 
\frac{\partial f^{\text{eq}}_{n\mathbf{k}}}{\partial \varepsilon_{n\mathbf{k}}} 
\right) 
\left\{ 
v^{n}_{a,\mathbf{k}} 
\left(  
\partial_{k_c} G^{nn}_{bd,\mathbf{k}} 
+  \partial_{k_d} G^{nn}_{bc,\mathbf{k}} 
- \partial_{k_b} G^{nn}_{cd,\mathbf{k}} 
\right) 
+ (a \leftrightarrow b) 
\right\}
+ \left( 
\frac{\partial^2 f^{\text{eq}}_{n\mathbf{k}}}{\partial \varepsilon_{n\mathbf{k}}^2} 
\right) 
G^{nn}_{cd,\mathbf{k}} v^{n}_{a,\mathbf{k}} v^{n}_{b,\mathbf{k}} \right] ~.
\end{equation}
\end{widetext}
Equation~\eqref{eq:bcp_noise} is the leading intrinsic contribution to thermal noise at quadratic order in the electric field for $\mathcal{PT}$-symmetric systems, and it is one of the main findings of our work. 
It comprises of two distinct terms: the first arises from field corrections to the band energy and wavefunctions, involving derivatives of the BCP. The second stems from the second derivative of the distribution function, reflecting the field-shifted occupancy in steady state. Both terms feature derivatives of the Fermi-Dirac function, underscoring that BCP-induced noise is predominantly a Fermi-surface phenomena, and it is finite only in metals.  
We emphasize that, this intrinsic component of noise can be finite in centrosymmetric systems with parity, nonmagnetic time-reversal symmetric systems, and in antiferromagnets with $ {\cal P} {\cal T} $ symmetry.

This quantum metric driven thermal noise provides a universal, disorder-independent probe of band geometry in systems where Berry-curvature effects are absent, complementing established responses like the nonlinear Hall effect~\cite{Liu_2022_3rd_Hall}. Specifically, in $\mathcal{PT}$-symmetric antiferromagnets (e.g., CuMnAs), it constitutes the dominant geometric term, potentially measurable via noise spectroscopy. When Berry curvature is nonzero, the BCP noise adds an additional contribution to the thermal noise, in addition to contributions from quantum-geometric-tensor fluctuations~\cite{Miaomiao_2023}. 
\section{Thermal noise in topological surface states} 
    To illustrate our findings, we examine the surface states of a magnetic topological material, which are described by a 
    tilted Dirac model with a nonzero bandgap. The low energy effective Hamiltonian of the gapped and tilted Dirac states 
    is described by, 
    \begin{equation} \label{Y_tilted_Dirac}
        H_{\mathbf{k}} 
        =  \hbar v_f  ( k_x \sigma_y - k_y \sigma_x ) 
        + \hbar \Delta \sigma_z 
        + \hbar v_t k_y \sigma_0 ~~.
    \end{equation} 
    Here, $v_f$ is the Fermi velocity, $\Delta$ is the gap parameter, and $v_t$ introduces a tilt along the $Y$ direction. $\sigma_i$ denote the set of three Pauli matrices with $\sigma_0$ being the $2\times2$ identity matrix. This model breaks both $ {\cal P}$  and ${\cal T}$ symmetry, and is simple enough to 
    analytically calculate the quantum metric. The band dispersion is given by 
    $ \varepsilon_{n, \mathbf{k}} = \hbar v_t k_y + s ( \hbar^2 v_f ^2 k^2 + \hbar^2 \Delta^2 )^{1/2} $, with $s = 1$ for the conduction band and $s=-1$ for the valence band, and $k = (k_x^2 + k_y^2)^{1/2}$. 
    The band velocity is given by, 
    \begin{equation} \label{Band_velocity_Y_tilted_Dirac}
        v_{ x, \mathbf{k} } ^{nn}
        = - s \frac{v_f ^2 k_x}{\Lambda} , 
        \quad
        v_{ y, \mathbf{k} } ^{nn}
        = v_t - s \frac{v_f ^2 k_y}{\Lambda} ~~.
    \end{equation}
    Here, we have defined $ \Lambda = ( v_f ^2 k^2 + \Delta^2 )^{1/2} $. The BCP tensor components are
    \begin{equation} \label{BCP_Y_tilted_Dirac}
        \begin{split}
            & G_{xx} ^{cc} 
            = \frac{v_f ^2}{ 4 \Lambda^5 } ( v_f ^2 k_y ^2 + \Delta^2 ) 
            = - G_{xx} ^{vv} \\
            & G_{yy} ^{cc} 
            = \frac{v_f ^2}{ 4 \Lambda^5 } ( v_f ^2 k_x ^2 + \Delta^2 ) 
            = - G_{yy} ^{vv} \\
            & G_{xy} ^{cc} 
            = - \frac{v_f ^4 k_x k_y}{ 4 \Lambda^5 } 
            = - G_{xy} ^{vv} ~~.
        \end{split}
    \end{equation}
    Since the tilt velocity only appears in the diagonal components of the Hamiltonian, it does not influence the eigenfunctions and consequently the BCP is independent of the $v_t$. However, $ v_t$ produces a constant drift in the band velocity along the direction of the tilt, and consequently influences the BCP-controlled noise. 

    In Figs.~\ref{fig:Dirac}(a)–\ref{fig:Dirac}(c), we present the BCP tensor components near $\mathbf{k}=(0,0)$, where the band gap is smallest. All BCP components exhibit a sharp peak at this point. This behaviour is directly reflected in the current correlations shown in Figs.~\ref{fig:Dirac}(d)–\ref{fig:Dirac}(f), all of which peak near the band edges. Both the longitudinal and transverse auto-correlations [\ref{fig:Dirac}(d),\ref{fig:Dirac}(e)] and their cross-correlation [\ref{fig:Dirac}(f)] are strongly enhanced when the chemical potential $\mu$ lies near the gap but vanish inside the gap, indicating their Fermi-surface origin. In all cases, the noise magnitude grows with temperature, consistent with the thermal factor $ f^{eq} _{n \mathbf{k}} ( 1-f^{eq} _{n \mathbf{k}} ) 
    = - k_B T ~(\partial f^{eq} _{n \mathbf{k}} / \partial \varepsilon_{n \mathbf{k}} )$ in equation~(\ref{eq:thermal_noise}). Next, we analyze the thermal noise in CuMnAs. 
    \begin{figure}[t!]
        \includegraphics[width = \linewidth]{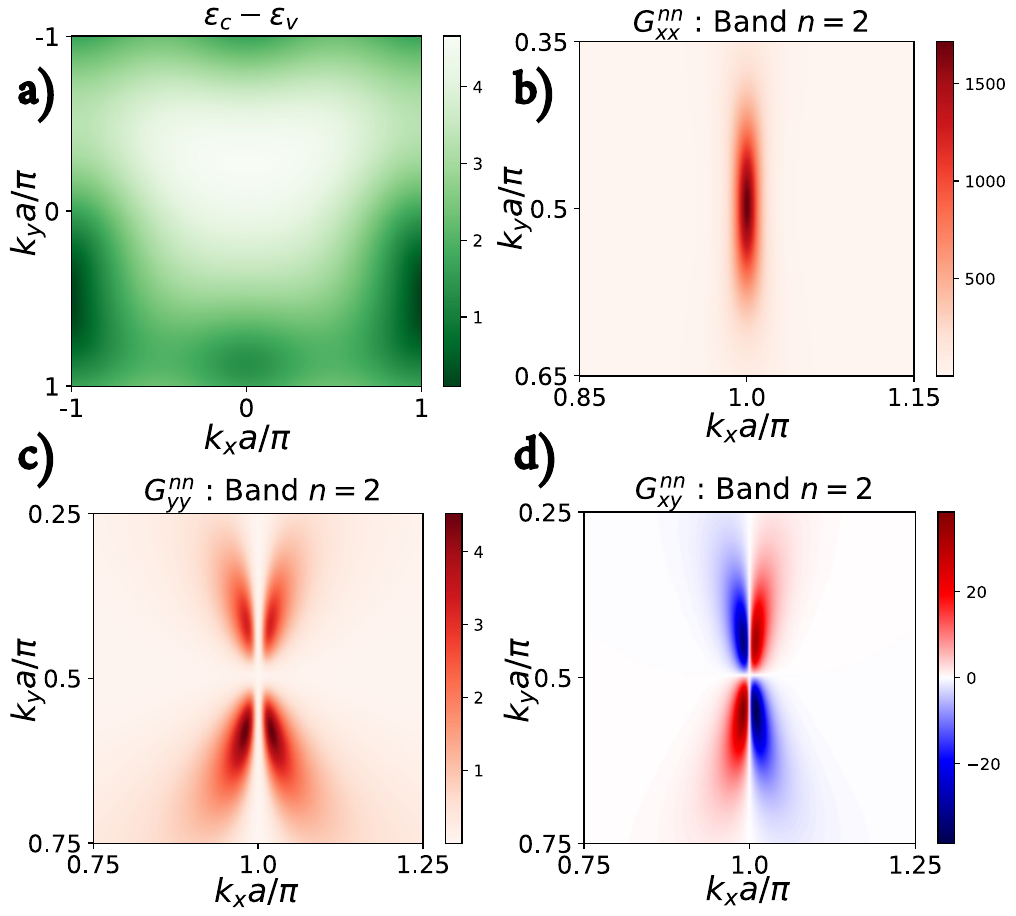}
        \caption[width=\columnwidth]{\textbf{CuMnAs band-gap and band-geometry:} (a) Momentum-space distribution of the energy gap between the conduction and valence band, with a minimum observed near $\mathbf{k} = (1,0.5)\pi$. (b)–(d) Conduction-band BCP tensor $G^{nn} _{ab}$ evaluated around this minimum-gap point, displaying their prominent features. We use the Hamiltonian in Eq.~\ref{CuMnAs_Hamiltonian} with parameters: $t_0=1$ eV, $t_1=0.08$ eV, $a_D=0$ eV, $a_R=0.8$ eV, and $(h_x,h_y,h_z)=(0.85,0,0)$ eV.
        \label{fig:CuMnAs_Band}}
    \end{figure}
     \begin{figure*}[t!]
        \includegraphics[width = .9\linewidth]{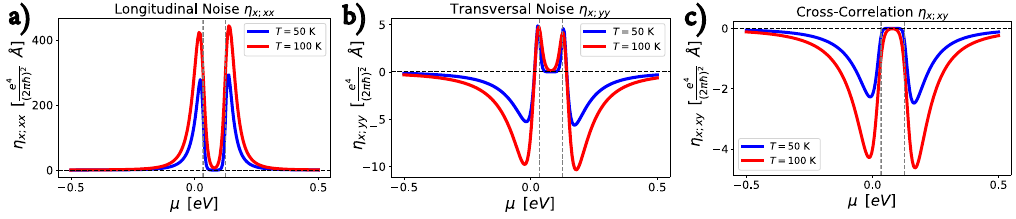}
        \caption{\textbf{CuMnAs Correlations:} Chemical potential dependence of different components of BCP-regulated intrinsic thermal noise. Panels (a) and (b) show thermal noise of currents along $X$ and $Y$, while (c) shows their cross-correlation. The electric field applied is $\mathbf{E}=(1,0,0)$ V/m, with $X$ and $Y$ aligned to the crystalline axis of CuMnAs. Inside the gap, the suppressed correlations illustrate the Fermi-surface effect, while the peaks near the gap-edges reflect the band-geometric origin. We used the Hamiltonian in Eq.~(\ref{CuMnAs_Hamiltonian}), with the same model parameters described in Fig.~\ref{fig:CuMnAs_Band}.
        \label{fig:CuMnAs_Correlations}}
    \end{figure*}
    \section{Thermal noise in ${\cal P}{\cal T}$ symmetric antiferromagnet CuMnAs} 
    Going beyond $\cal T$ and $\cal P$ broken system, we next study thermal noise in CuMnAs, a ${\cal P}{\cal T}$-symmetric antiferromagnet. Such antiferromagnetic systems provide an ideal platform to highlight nonlinear effects driven by the quantum-metric or BCP, since competing effects due to Berry curvature are forbidden by symmetry. 
    
    CuMnAs hosts anti-parallel spins on a bipartite lattice of sublattices $A$ and $B$. This arrangement breaks both the parity and the time-reversal symmetries. However, exchanging the sublattices (${\cal P}$ operation) and flipping the spins (${\cal T}$ operation) restores the system to its original configuration, thereby preserving the combined ${\cal P}{\cal T}$ symmetry. The effective physics of CuMnAs can be described by an effective tight-binding model in two dimensions\cite{Tang_2016_Dirac_CuMnAs, 
    Smejkal_17_CuMnAs, 
    Watanabe_21_CuMnAs_Chiral}. The Bloch Hamiltonian in the sublattice basis is given by, 
    \begin{equation} \label{CuMnAs_Hamiltonian}
        H_{ \mathbf{k} } = 
        \begin{pmatrix}
            \epsilon_0 ( \mathbf{k} ) ~\sigma_0 
            + \mathbf{h}_A ( \mathbf{k} ) 
            \cdot \boldsymbol{\sigma} 
            & V_{AB} ( \mathbf{k} )  ~\mathbf{1}_2 \\
            & \\
            & \\
             V_{BA} ( \mathbf{k} )  ~\mathbf{1}_2 
            & \epsilon_0 ( \mathbf{k} ) ~\sigma_0 
            + \mathbf{h}_B ( \mathbf{k} ) 
            \cdot \boldsymbol{\sigma}
        \end{pmatrix} \enspace ~~.
    \end{equation} \\
    In the above equation, the nearest-neighbor hopping is encoded in the inter-sublattice ($A$-$B$) term as
    $ V_{AB} ( \mathbf{k} ) 
    = - 2 t_0 \cos( k_x / 2 ) \cos( k_y / 2 ) 
    = V_{BA} ( \mathbf{k} ) $, where $t_0$ is the hopping parameter between the nearest atoms. 
    In the diagonal term, $ \epsilon_0 ( \mathbf{k} ) 
    = - t_1 ( \cos k_x 
    + \cos k_y  ) $ captures the inter-cell hopping between the same sublattices ($A$-$A$ or $B$-$B$), characterised by $t_1$, the next-nearest neighbor hopping strength.
    The diagonal elements further incorporate the sublattice-dependent coupling with the electronic spin ($\bm{\sigma}$), given by 
    $ \mathbf{h}_A ( \mathbf{k} )
    = - \mathbf{h}_B ( \mathbf{k} ) 
    = (  h_x 
    - a_R \sin k_y 
    + a_D \sin k_y , 
    \enspace h_y 
    + a_R \sin k_x 
    + a_D \sin k_x ,  
    \enspace h_z ) $. Here, $ ( h_x, h_y, h_z ) $ represents the antiferromagnetic magnetization field, while $ a_D, a_R $ denote the Dresselhaus and Rashba spin-orbit coupling strengths, respectively. The Hamiltonian admits four bands, but ${\cal P}{\cal T}$ symmetry enforces a twofold degeneracy within both the valence and conduction bands. 

    We present the BCP tensor and the resulting BCP-induced noise in CuMnAs in Figs.~\ref{fig:CuMnAs_Band}, and \ref{fig:CuMnAs_Correlations}. Figure~\ref{fig:CuMnAs_Band}(a) shows the band gap across the Brillouin zone, with one of the Dirac points being at $\mathbf{k} = (1,0.5)\pi$. The conduction-band BCP components are peaked about these Dirac points, as seen in Figs.~\ref{fig:CuMnAs_Band}(b)–(d). Using the same model parameters, we evaluate the noise spectrum for an applied field $\mathbf{E}=(1,0,0)$ V/m in Fig.~\ref{fig:CuMnAs_Correlations}. We find that the longitudinal, transverse, and cross-correlated noise components vanish inside the band gap but display sharp peaks at the band edges. This is consistent with their Fermi-surface origin and the enhancement of the BCP near the band edges.
\section{Discussion}
The formalism developed here establishes thermal noise as a versatile probe of quantum geometry in nonlinear transport, extending beyond Berry-curvature–dominated responses to metric-driven noise. In ${\cal PT}$-symmetric antiferromagnets, where Berry curvature vanishes across the Brillouin zone, the BCP emerges as the principal band-geometric contributor to second-order intrinsic noise. This $\tau$-independent contribution originates from field-induced renormalizations of Bloch wavefunctions and band energies, and complements the extrinsic Drude term that scales as $\tau^2$.

A limitation of our approach is the use of a constant relaxation time. More realistic treatments could incorporate static disorder (point defects, charged impurities) and dynamic scattering (phonons), with the latter being essential for capturing the temperature dependence of thermal noise. These refinements, however, do not affect the intrinsic contribution. Another natural extension is to include asymmetric scattering contributions to the thermal noise arising from the side-jump and skew scattering processes.

Our work highlights the role of quantum geometry in fluctuation phenomena. In antiferromagnetic spintronics, where stray fields are restricted, BCP-driven noise could serve as a direct experimental probe of hidden topological phases that remain inaccessible to conventional transport measurements.
    \section{Conclusion} 
    We have shown that second-order thermal noise in charge currents contains an intrinsic contribution governed by the quantum metric. This contribution is encoded in the Berry connection polarizability (BCP), reinforcing its central role beyond nonlinear charge transport. We illustrated these effects in Dirac surface states and highlighted their significance in $\mathcal{PT}$-symmetric antiferromagnets, where Berry curvature vanishes identically. Our results establish second-order BCP thermal noise as a symmetry-protected intrinsic probe of band geometry in quantum materials.
    \section{Acknowledgement}
    We thank Dr. Kamal Das, Dr. Debottam Mandal and Harsh Varshney for insightful discussion. D. Bhowmick is supported by the Institute Fellowship, IIT Kanpur. A. Agarwal acknowledges funding from the Core Research Grant by ANRF (Sanction No. CRG/2023/007003), Department of Science and Technology, India.

    \bibliography{Bibliography}

\end{document}